\def\etal{{\it et al.\thinspace}}

\def\ie{{\it i.e.\ }}

\def\gcm3{{g cm${}^{-3}$}}

\def\h50{\hbox{$\rm\thinspace h_{50}$}}
\def\h50m1{\hbox{$\rm\thinspace h_{50}^{-1}$}}

\def\etal{{\it et al.\thinspace}}

\def\ie{{\it i.e.\ }}

\def\Fig{Figure}

\def\p3m{P${}^3$M}
\def\ap3m{AP${}^3$M}
\def\-{{\em{---}}}

\documentstyle[emulateapj,psfig,apjfonts]{article}

\received{}
\accepted{}
\journalid{}{}
\articleid{}{} 

\lefthead{Barnes \& Raymond}
\righthead{Predicting Planets}

\begin{document}

\title{Predicting Planets in Known Extra-Solar Planetary Systems I: Test Particle Simulations}

\author{Rory Barnes\altaffilmark{1}, Sean N. Raymond\altaffilmark{1}}
\altaffiltext{1}{Department of Astronomy, University of Washington, Seattle, 
WA, 98195-1580}

\begin{center}E-mail:rory@astro.washington.edu
\end{center}

\begin{abstract} Recent work has suggested that many planetary systems lie
near instability. If all systems are near instability, an additional planet
must exist in stable regions of well-separated extra-solar planetary systems
to push these systems to the edge of stability. We examine the known systems
by placing massless test particles in between the planets and integrating for
1-10 million years. We find that some systems, HD168443 and HD74156, eject
nearly all test particles within 2 million years. However we find that
HD37124, HD38529, and 55Cnc have large contiguous regions in which particles
survive for 10 million years. These three systems, therefore, seem the most
likely candidates for additional companions. Furthermore HD74156 and HD168443
must be complete and therefore radial velocity surveys should only focus on
detecting more distant companions. We also find that several systems show
stable regions that only exist at nonzero eccentricities.

 \end{abstract}



\section{Introduction}

Several categories of planetary systems have been discovered in the
past several years. Some appear dynamically similar to our own Solar
System (SS), but others appear quite different (Barnes \& Quinn 2004,
hereafter BQ). Whether these systems really fall into
unique categories, or are the first examples of a continuous
spectrum of stability remains to be seen. Nonetheless we will define three categories of
systems: resonant, interacting, and separated. For the purposes of this paper, we will define a planetary system as a star with 2 or more companions; at least a 3-body system. Resonant systems contain 2 or more planets in mean motion
orbits. The GJ876 (Butler \etal 2001) and
HD82943\footnote{http://www.obs-geneve.ch/$\sim$udry/hd82943syst.html}
systems are in 2:1 resonance, and the 2 inner planets of the 55Cnc system (Marcy \etal 2002)
are in 3:1 resonance. The interacting systems of $\upsilon$ And (Butler
\etal 1999), HD12661 (Fischer \etal 2003), 47UMa (Butler \etal 2001),
and the SS are not in resonance, but their orbits are close enough that the planets may perturb each other. The
final category is the separated systems. In these systems, the planets
are separated enough that they are not interacting on long timescales
($\lesssim 10^6$ years). This paper will focus on these separated systems, 
specifically  HD168443 (Marcy \etal 2001), 
HD74156\footnote{http://obswww.unige.ch/$\sim$udry/planet/hd37124.html}, 
HD38529
(Fischer \etal 2003), and HD37124 (Butler \etal 2003). We will also examine 
55Cnc, which is a combination of a resonant and a separated system.

Studies of the known resonant and interacting systems have demonstrated that
these systems are on the verge of instability (BQ). Slight changes in
orbital elements, specifically eccentricity or proximity to perfect
resonance, lead to a catastrophic disruption of the systems. Planetary
systems near instability are as tightly packed as
possible; there is no room for additional companions. In this paper we
explore the possibility that all systems are tightly packed, implying that
additional companions lie between the extant planets in separated systems.
In this ``packed planetary systems'' (PPS) hypothesis, the undetected planets have not been observed because there are not
enough data to discover the additional planet, or the planetary mass falls
at or below the detection limit of current Doppler technology.  BQ also
examined the stability of the Sun-Jupiter-Saturn system and showed that
it lies further from instability than the complete gas giant system. The Sun-Jupiter-Saturn model is the only system
known to be incomplete and therefore supports the hypothesis that all systems lie near instability, and are hence packed.

The PPS model assumes that planet
formation is an efficient process. As many planets (or at least gas
giants) form in a circumstellar disks as possible. This is seen 
in the gas giant region of the SS, as
well as in the resonant and coupled systems. In this scenario, as dust congeals into ever larger bodies,
the mutual gravitational forces between protoplanets perturb each
other in ever increasing amounts as the masses grow. As the protoplanets
become larger, they acquire the ability to pull other planets into
favorable positions (dynamically stable), or, if the perturbations
become strong enough, eject them completely from the system. The
implication is that any stable region in between known planets
harbors an (as yet) unseen companion. In this paper we attempt to map
out stable regions in semi-major axis and eccentricity space of known
separated systems.

We perform numerical simulations using test (massless) particles to
search for stable regions in HD168443, HD74156, HD37124, HD38529, and
55Cnc. In addition to these systems we also perform a control
experiment. We replace the middle planet of the $\upsilon$ And
system with a belt of test particles to evaluate how well this experiment predicts the orbit $\upsilon$
And c. We find that some systems, HD74156 and HD168443, contain no
stable regions and are complete to the most distant planet. HD37124,
HD38529 and the resonant/separated system 55Cnc contain a zone of
stability which may harbor an undetected planet. This paper is divided
into the following sections. In $\S$2 we describe the numerical
techniques used to examine these systems. In $\S$3 we present the
results for systems which may have additional companions. In $\S$4
we present the results of the $\upsilon$ And system, with the middle
planet removed. We summarize our results in $\S$5 and suggest future
work in this field. This is the first in a series of papers which will
explore the possibility of additional planets in known systems. In
Paper II we will place massive planets in the stable regions
identified by this work to refine our predictions. In Paper III we
examine how planet formation might occur between the
detected planets.

\section{Numerical Methods}

These simulations were performed with SWIFT\footnote{SWIFT is publicly
available at http://www.boulder.swri.edu/$\sim$hal/swift.html} (Levison \& 
Duncan
1994). Specifically we used the regularized mixed variable symplectic
integrator, RMVS3. This code is designed to quickly
integrate a system through close approaches (via regularization), yet
still maintain conservation of energy (via a symplectic
algorithm). For most systems (except 55Cnc and $\upsilon$ And) we
integrate for $10^7$ years. We require every simulation to conserve
energy to a factor of $10^{-4}$, which has been shown to be accurate
enough for simulations of planetary system stability on these
timescales (BQ). Generally we conserve energy to better than 1 part in
$10^6$.

The systems presented here are all coplanar. The mass-inclination
degeneracy of Doppler observations is broken by assigning the
companions' masses to be the observed minimum masses. All other
orbital elements are their best fit values. Note that in
coplanar systems, the longitude of ascending node is meaningless. Test
particles are all also coplanar. They are spaced every 0.002AU in
semi-major axis space, and every 0.05 in eccentricity space. Their
mean anomalies are placed uniformly throughout [0,2$\pi$). In
non-circular trials the longitude of periastron of the test particles was 
aligned with the
most massive planet. This alignment is arbitrary, but probably valid
for low e ($<0.1$). However some research suggests that for $e>0.1$ the 
apses should have been
anti-aligned (Laughlin, Chambers \& Fischer
2002), although this phenomenon is not strictly necessary (BQ). 

This 
procedure is similar to other work on test particles
in $\upsilon$ And (Rivera \& Lissauer 2000), GJ876 (Rivera \& Lissauer
2001), and 55Cnc (Rivera \& Haghighipour 2002). The major difference
being that we examine non-circular test particle orbits. Additionally Menou \& Tabachnik (2003) placed test particles in the habitable zones of these systems. With the exception of HD37124, we recover their results.

\section{Results}
In Table 1, the orbital elements of all the planetary systems used in
this paper are shown. As there are no published data for some systems,
we are forced to use data from the internet, which are constantly
changing.  In $\S$3.1 we examine the complete systems of HD74156 and HD168443 and in $\S$3.2 we analyze the HD37124, HD38529, and 55Cnc systems which contain zones of stability.

\begin{figure*}
\psfig{file=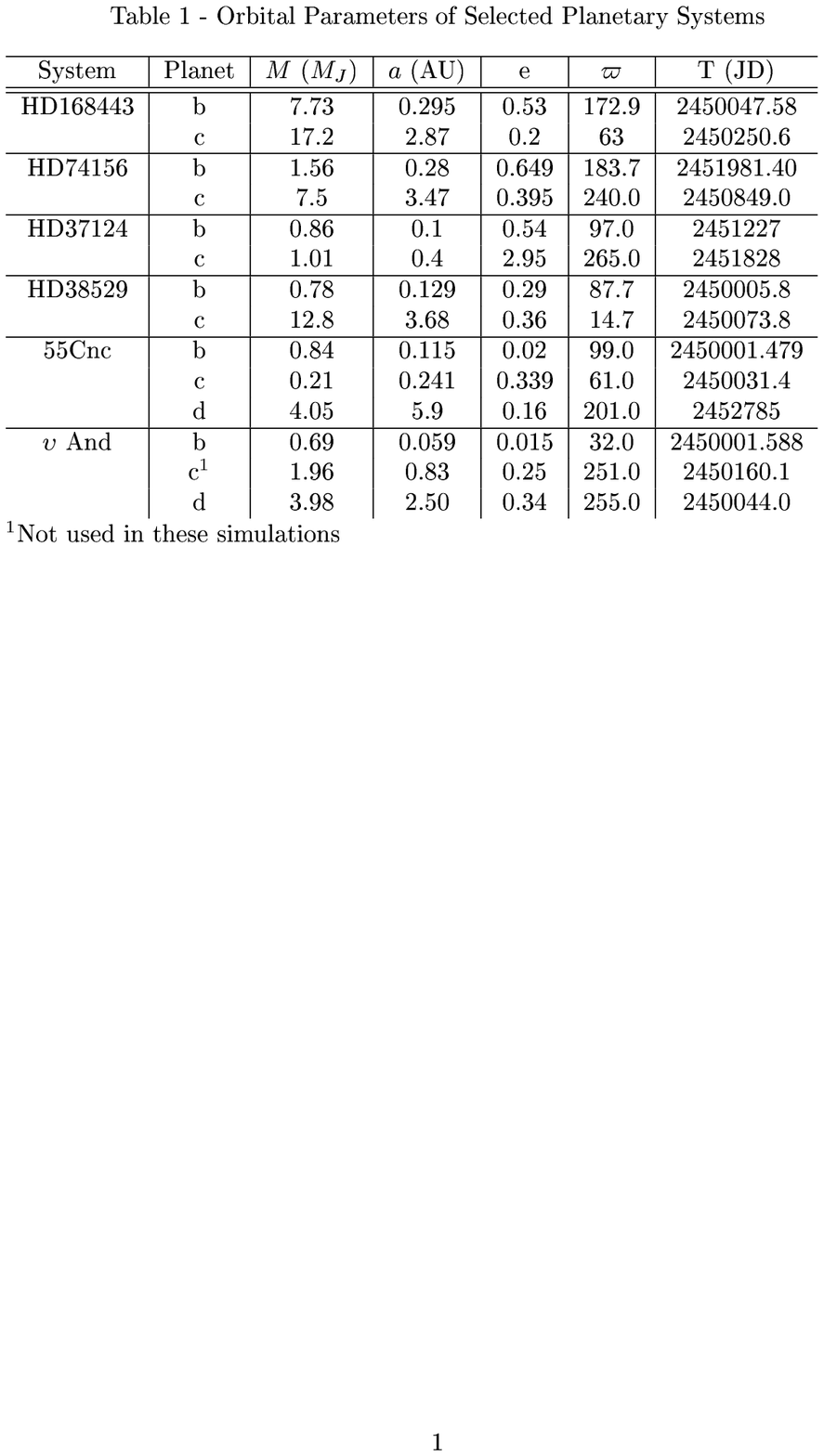,width=19.truecm}
\end{figure*}

\subsection{Complete Systems: HD74156 \& HD168443}
From Table 1 nothing about these 2 systems would indicate \textit{a priori} that they would have no stable
zones, although their planetary masses are some of the largest of this
subset of systems. Note, however, that HD38529c is the 2nd largest
planet examined, and it has one of the largest zones of stability (see
$\S$3.2). The ratios of the periods, $R$, of HD74156 and HD168443 are
51.4 and 30.5, respectively, which are not the smallest values among
separated systems. Nonetheless these systems show little evidence for stable particles between the currently known planets.

In HD168443, $R$=30.5, and we need 786 test particles to fill the
region between planets b and c. In Figure 1, we see that regardless of
eccentricity or semi-major axis no test particles survive for even 2 
million years. We therefore conclude that this
system is complete out to planet c, there can be no asteroid belt in this
system, and radial/astrometric surveys should sample this system sparsely.

\medskip
\epsfxsize=8truecm
\epsfbox{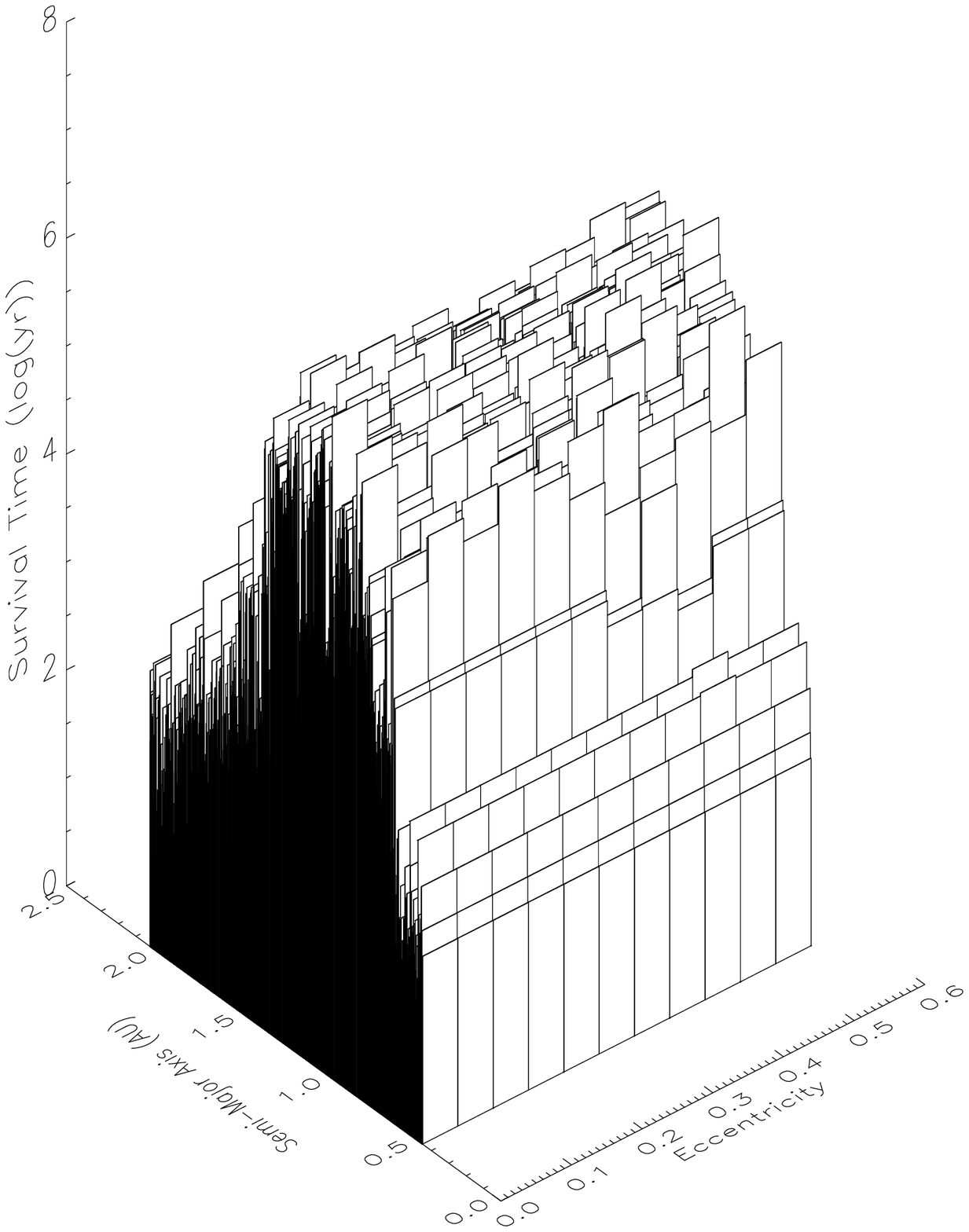}
\figcaption[f1.eps]{\label{fig:asymptotic}
\small{Stability of test particles in HD168443. The height of the bars 
corresponds to how long test particles remained bound to the parent star. 
Regardless of eccentricity, no test particles can survive in this system 
for even 2 million years.}}
\medskip

The best fit orbital elements for HD74156 have fluctuated throughout the past
several years. The elements presented in Table 1 are from the
discovery website and date to May of 2002. The elements changed
dramatically in August of 2002 (namely $a_c$ increased to 3.8AU). The
data have recently been published (Naef \etal 2003), and the elements
changed to values similar to those of May 2002. Although the system
presented here is slightly different than the current best
fit, our work is similar enough that any differences should be
negligible.

HD74156 consists of two planets separated by nearly a factor of 40 in
period. We need 822 test particles for this system. The
eccentricities and masses are lower than in HD168443. In Figure 2, we
present the survivability of test particles in this system. A very
narrow band of stability may exist at 0.6AU, this strip is likely
unstable on timescales larger than $10^7$ years. Note that this strip
only develops at $e>0.1$. Therefore any asteroids that may exist in
this system must be on significantly eccentric orbits. Although it
appears unlikely that any planet could exist in the putative region,
we will show in Paper II (Raymond \& Barnes, in preparation) that
Saturn mass companions can survive in the system for at least $10^8$yr.

\medskip
\epsfxsize=8truecm
\epsfbox{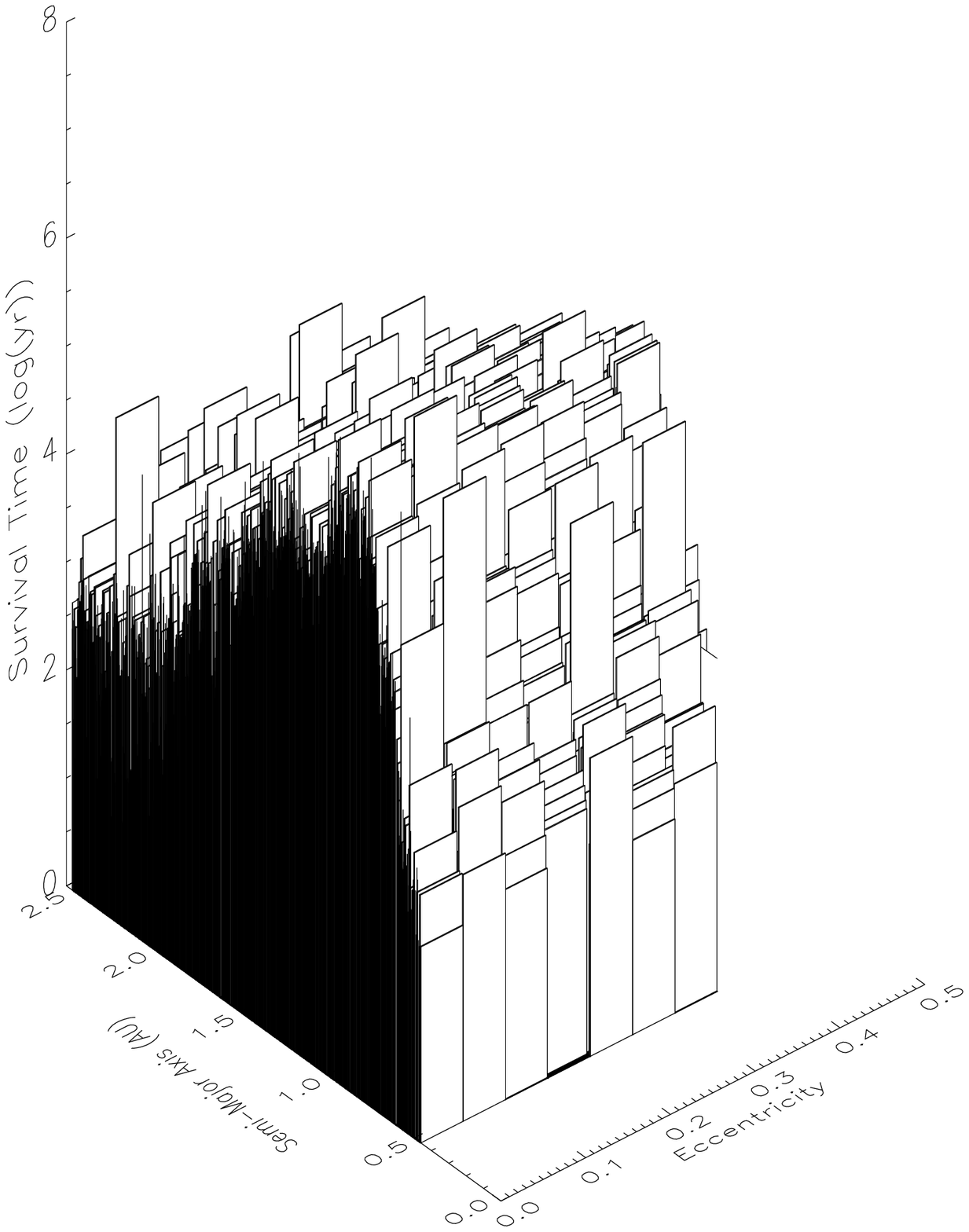}
\figcaption[f2ps]{\label{fig:asymptotic}
\small{Stability of test particles in HD74156. The stable region
located at $a=0.6AU, 0.1<e<0.25$ is so narrow that an additional
companion is very unlikely. Note that the band arises
at nonzero eccentricity.}}
\medskip

\subsection{Candidate Systems: HD37124, HD38529, \& 55Cnc} 
Three systems show broad regions of stability for test particles. The
current orbital parameters of HD37124, HD38529, and 55Cnc are presented in
Table 1. HD37124 and HD38529 are classical separated system, whereas
55Cnc is a resonant system with a distant, separated companion. In 55Cnc
the outer two planets are separated by $R$=121, HD37124, $R$=10.1, and
HD38529, $R$=152.

In Figure 3, we show the stable zone for HD37124. For this
system $R$=12.7 and we integrated 588 test particles. Although dotted
with ejections due to high order resonances, there do appear to be
significant regions of stability, most notably is at a semi-major axis of 
approximately 1AU. There is also an increase in  stability near $0.15\le e\le 0.2$. As in 
previous systems, we note that the stable region is 
not largest at e=0. Therefore we suggest that the most likely orbit for an
additional companion is at a semi-major axis slightly smaller or
larger than 1AU (which corresponds to the 5:2 resonance with planet
b), and an eccentricity near 0.15. Should there be no planet here,
these results suggest that the presence of an asteroid belt in this
region of phase space is likely.

\medskip
\epsfxsize=8truecm
\epsfbox{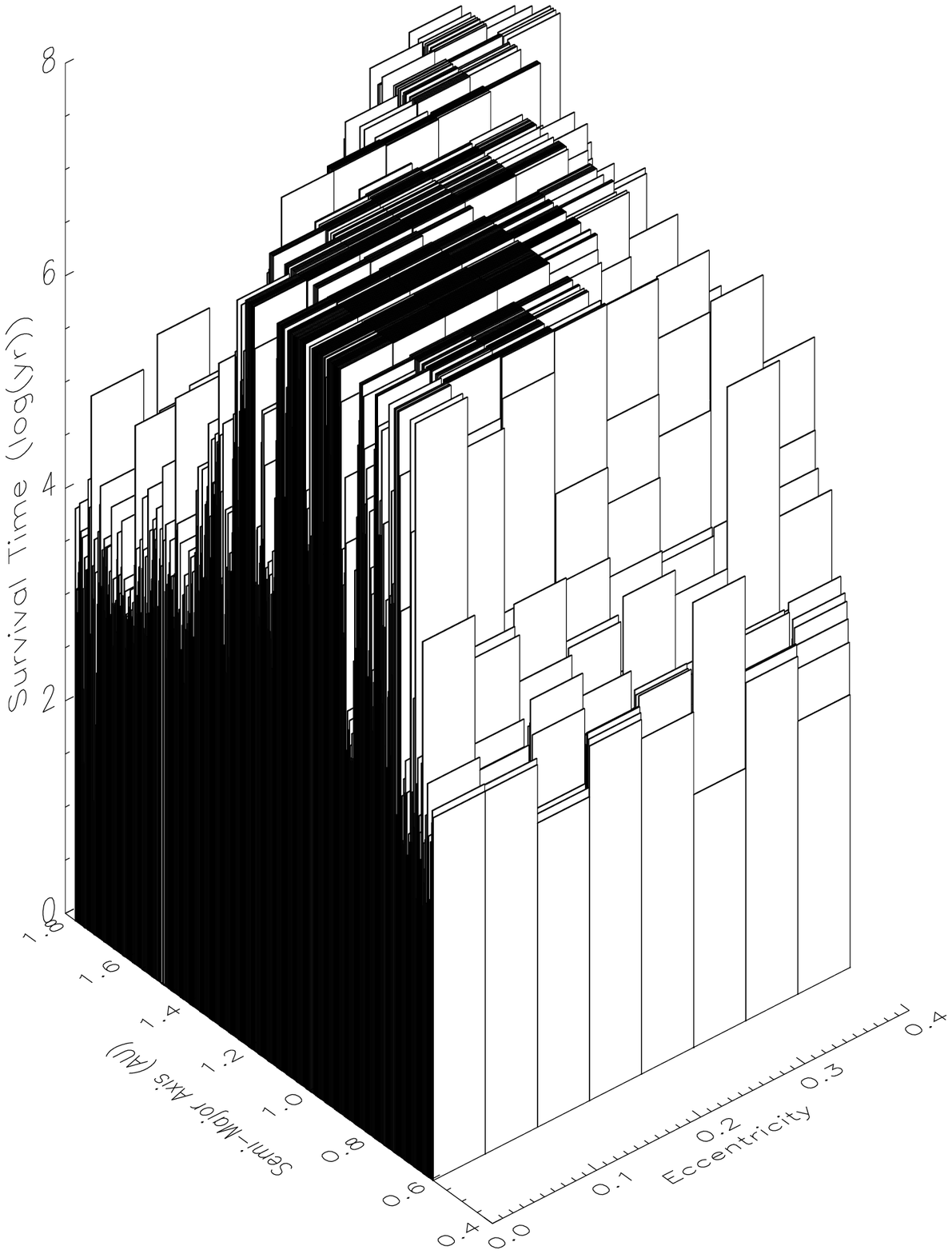}
\figcaption[f3.eps]{\label{fig:asymptotic}
\small{Stability of test particles in HD37124. The most likely orbit for an additional companion lies just interior or exterior to 1AU, with an eccentricity of $\sim 0.15$. This plots also shows the possible orbits of asteroids should no additional planet be present.}}
\medskip

 Note also in Fig.\ 3 that an
additional band of stability arises at $a\approx 1.2$AU above e=0.1. A
third band is visible at $a=0.9$AU, $e=0.15$. A fourth band arises at
$a\approx 1.7$AU at $e>0.3$. This suggests that the significant
(\ie larger than 0.1) values of extra-solar planetary eccentricities
may not be detrimental to system stability. This has been suggested
before (Murray \& Dermott 1999, Menou \& Tabachnik 20003), but is
shown dramatically in Fig.\ 3. Although moderate eccentricities may encourage
stability, at some point in every system, a critical
eccentricity is reached and the system becomes unstable. For HD37124
this threshold is at e=0.3.

As mentioned in $\S$2, Menou \& Tabachnik obtained a different result
for this system. They integrated the region from 0.6 to 1.2AU for
$10^6$ years, with test particles separated by 0.006AU. They found
that no test particles, with eccentricities close to 0.05, survived in
this region. The discrepancy results from their stringent criteria for
ejection, such as labeling test particles which cross the boundary of
the habitable zone as unstable. We find test particles in our
habitable zone (in semi-major axis space) often have significant
eccentricities, and most likely crossed the boundary at some point
during the simulation.

In Figure 4 the stable regions of HD38529 are shown. $R$ for this
system is 152, and hence we integrated 1092 test particles. This
region also contains a narrow zone of stability. In this system we see
that stability lies in between 0.25 and 0.75AU, and $e<0.3$. The interior
edge of this stable zone is quite sharp, probably because the mass of
the inner planet is 15 times smaller than the outer. This system has a
larger zone of stability than HD37124, so it seems more likely that a
planet might exist in this system.

\medskip
\epsfxsize=8truecm
\epsfbox{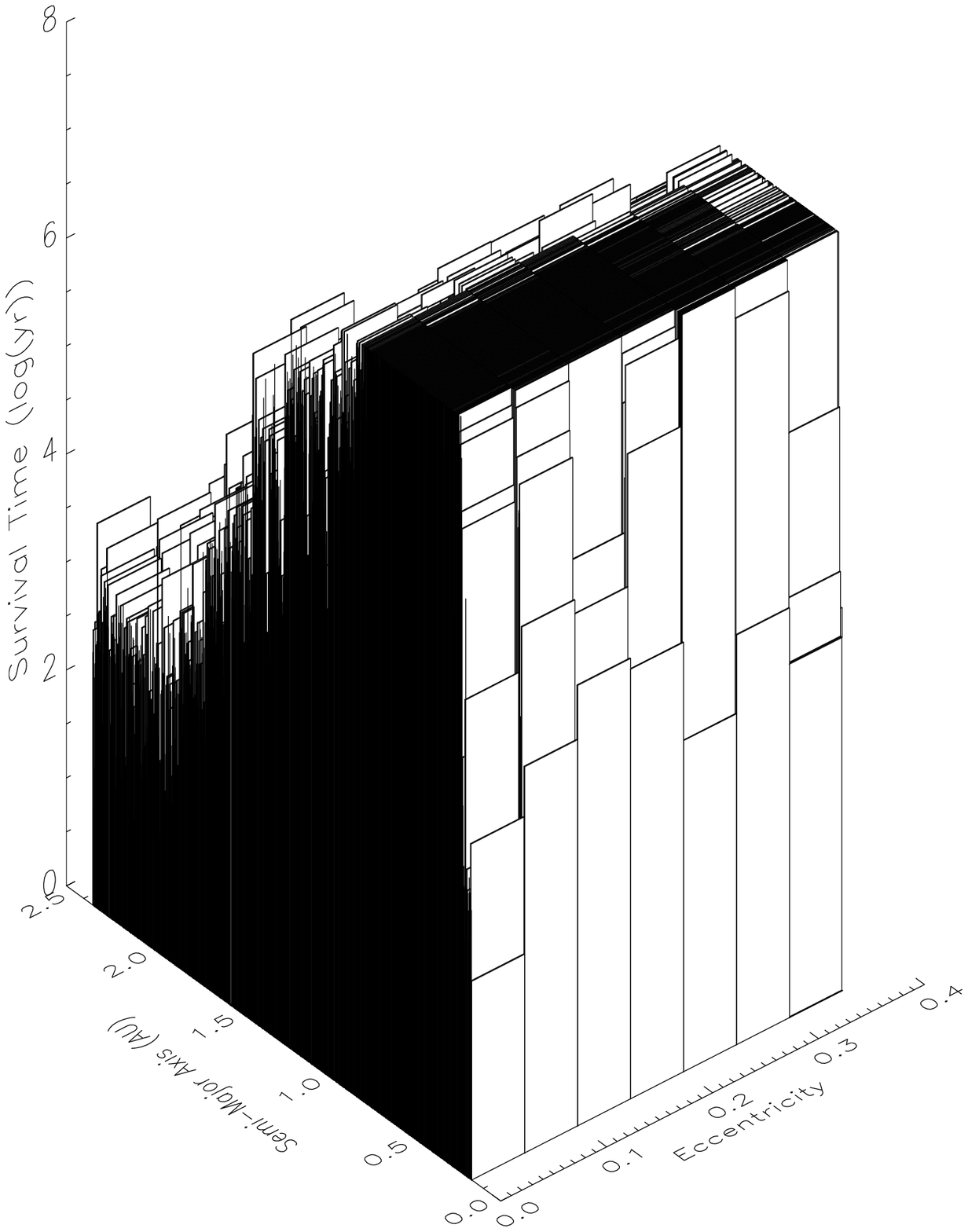}
\figcaption[f4.eps]{\label{fig:asymptotic}
\small{Stability of test particles in HD38529. Stability is most likely in the region $0.25AU\lesssim a\lesssim 0.75AU$. Above $e=0.3$ the probability of survival decreases dramatically.}}
\medskip

55Cnc is the only system with a
resonant pair, and a distant companion. Of all the systems examined
here, $R$ is the largest. Because of this wide separation, 55Cnc requires
2164 test particles and we could only integrate the system for 5
million years due to resource constraints. In Figure 5 we plot survivability in this system. Not
surprisingly, this system shows the broadest range of stability for
test particles. For all eccentricities there appears to be stability
from 0.9AU to 2.8AU. From 2.8AU to 3.5AU the stable regions are broken
up by mean motion resonances with planet d. There does appear to be a
slight preference for low eccentricities in this system. Therefore we
suggest that a planet might exist close to 2AU (the center of the
stability zone), and with $e\lesssim 0.1$. As before we also suggest
that this region might harbor an asteroid belt or terrestrial planets if it does not contain a gas giant.

55Cnc was also examined by Rivera \& Haghighipour (2002). They
examined test particles between planets c and d for 5 million years,
but all the test particles had zero eccentricity. They find a similar
stable region. This is also consistent with the results of Marcy \etal
(2002) which stated that an additional planet at 1AU would be
stable. Rivera \& Haghighipour also examined the region exterior to
planet d and found stability did not appear again until beyond
10AU. Given the similarity between our results and theirs, it is
likely that we would obtain a similar result, even for nonzero
eccentricities.

\medskip
\epsfxsize=8truecm
\epsfbox{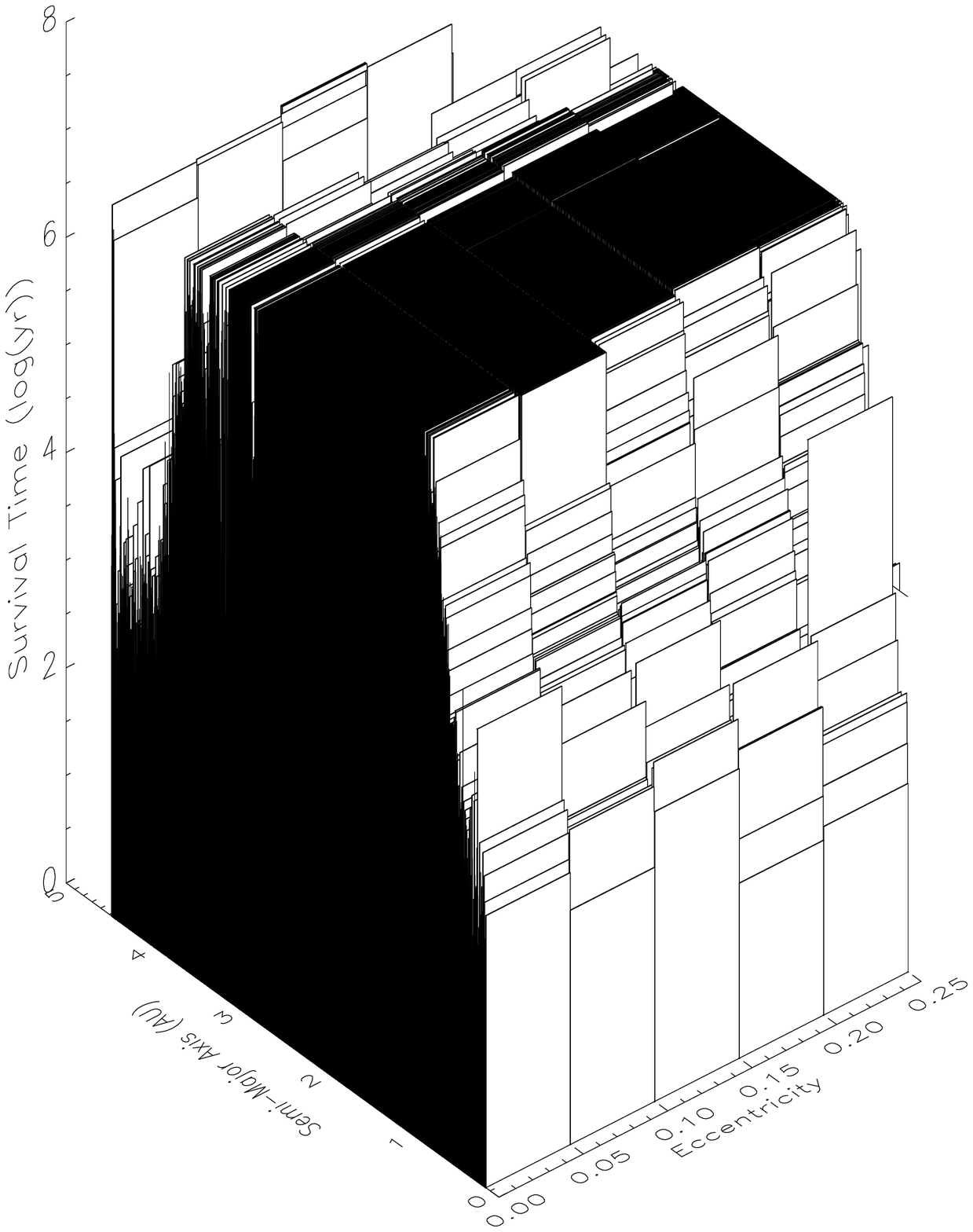}
\figcaption[f5.eps]{\label{fig:asymptotic}
\small{Stability of test particles in 55Cnc. A 1.9AU wide region of stability exists in this system. Unlike other systems, high eccentricity does not appear to promote stability. This system seems the best candidate for additional companions.}}
\medskip

Although there is a broad region of stability, this is only true of
massless particles. The resonant pair in this system lies close to
instability (Henderson \& Barnes, in preparation), therefore an
additional perturbing mass may increase the eccentricity, or move the
system from perfect resonance, and hence destabilize the system. In
Paper II we will address this issue when we place massive companions in
this region.

\section{A Control: $\upsilon$ Andromedae}
To test our methodology we simulated $\upsilon$ And to
determine if we could predict where the middle planet (planet c) might
lie. In Table 1 we presented the orbital elements of the planets in this
system. In our 3 body model, $R$=284 and we required 814 test
particles. The region between planets b and d proved so stable that
we were only able to integrate this system for 1 million years due to CPU time limitations. The
stability of this system is shown in Figure 6.  As we can see from this figure, we do indeed find
that planet c lies in the stable region.

\medskip
\epsfxsize=8truecm
\epsfbox{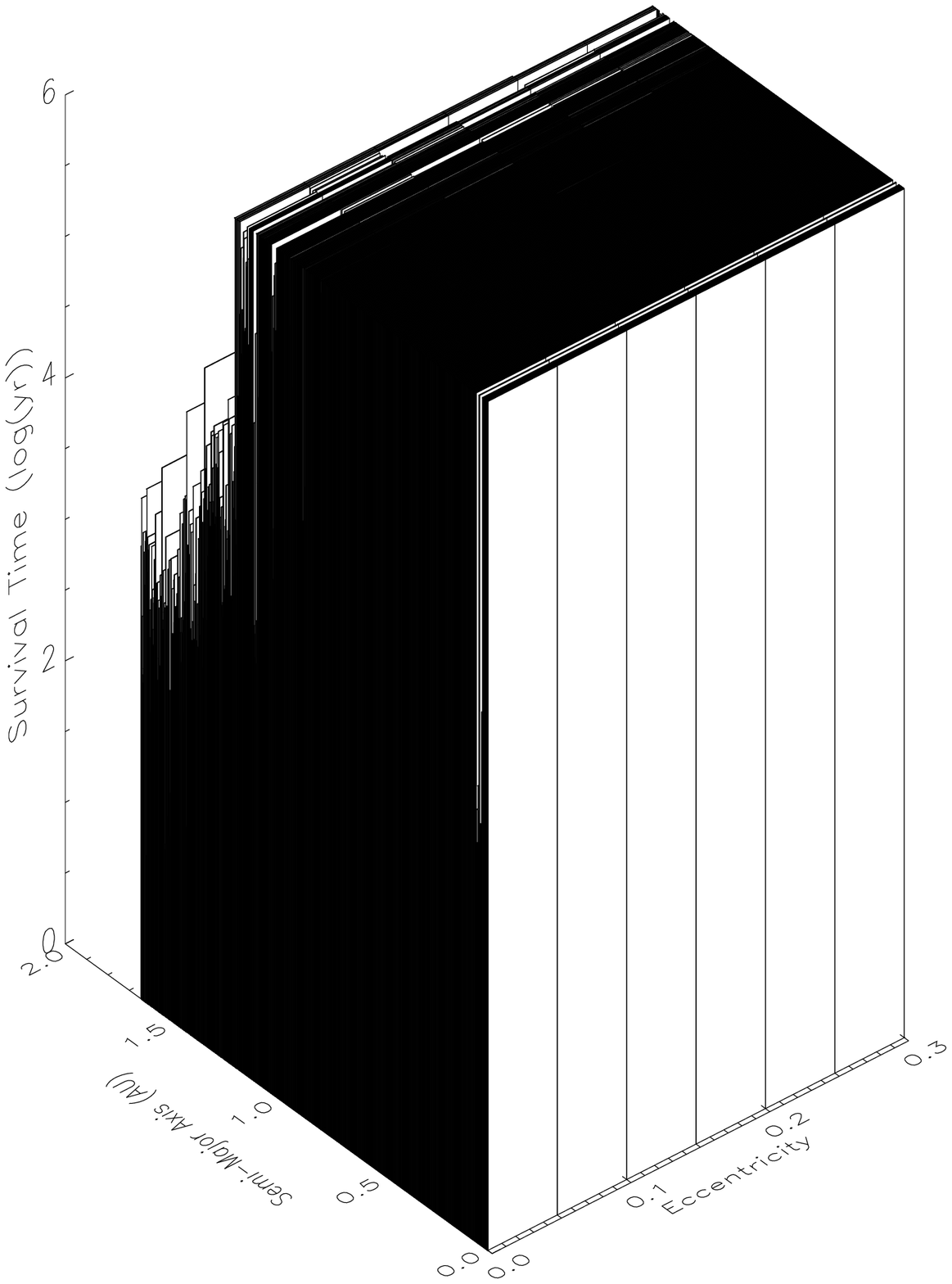}
\figcaption[f6.eps]{\label{fig:asymptotic}
\small{Stability of test particles in $\upsilon$ And. This system shows stability between 0.1 and 1.2AU. Note also that the system is more stable at eccentricities larger than 0.2. The current values of $a_c$ and $e_c$ are 0.8AU and 0.24, respectively.}}
\medskip

Test particle simulations of $\upsilon$ And were also performed by
Rivera \& Lissauer (2000). Their simulations ran for 5 million
years. They used all three planets, but did place test particles
between planets b and c, in addition to particle beyond the orbit of
planet d. They found that a very narrow region of stability between
planets b and c of approximately 0.35AU in width. They, too, find a
sharp edge near planet b. Our hypothesis should therefore suggest that
an additional planet may lie in this region between planets b and c as well (see $\S$5). However this is
the narrowest zone of stability seen in any system, and, given that
$\upsilon$ And already lies near the edge of stability, an additional
planet located here may disrupt the system.

\section{Discussion and Conclusions}

Building on the work of BQ we have argued that additional planets
should exist in separated systems so that they, too, lie close to
instability. We have tested the PPS hypothesis by integrating the orbits of a
large number of massless test particles in five known extra-solar planetary
systems. The results of these simulations are summarized in Table 2. In this table $N$ is the number of test particles used, $\Delta a$ and $\Delta e$ are the approximate zones of stability where an additional companion might exist. We find that some systems cannot contain
additional planets (HD74156 and HD168443), while others have significant stable
zones (HD37124, HD38529, and 55Cnc). From our control experiment, we
see that \Fig 6 most resembles \Fig 5, again suggesting that 55Cnc is
the most likely candidate for an additional companion.

\begin{figure*}
\psfig{file=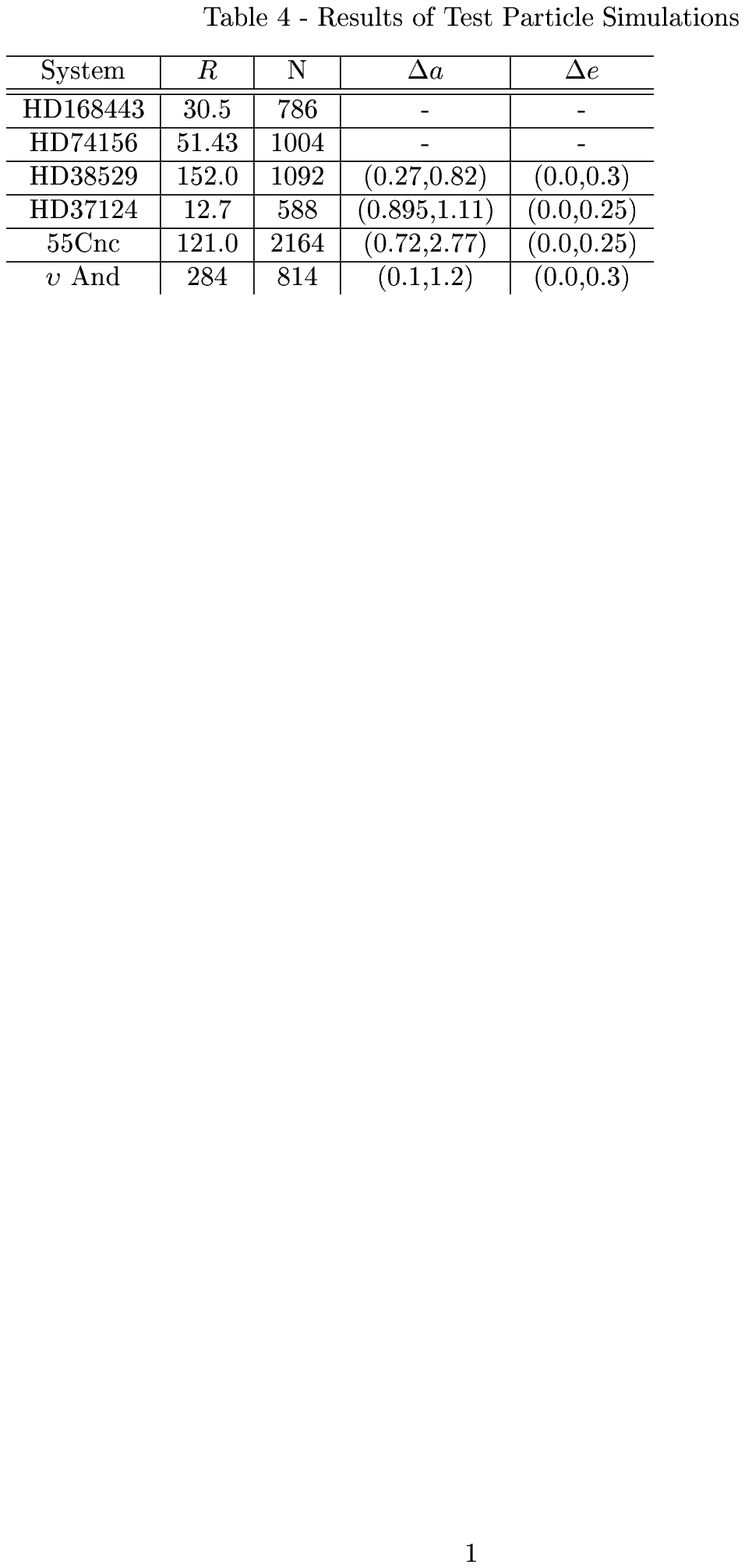,width=19.truecm}
\end{figure*}

We must note one ambiguity in the PPS hypothesis. As has been shown in other work
(Barnes \& Quinn 2001, BQ), $\upsilon$ And is already near the
edge of stability. Perhaps an additional companion was in this system, but it was
dynamically unstable, and ejected. This left c and d on
interacting orbits, and a small annulus of stability.  Planetary
systems with any 2 planets on the edge of stability may negate the
possibility of predicting additional companions. They are already on
the edge, and hence need no additional companions to push them
there. This is especially relevant in 55Cnc. The resonant pair is on
the edge (Henderson \& Barnes, in preparation), so perhaps the large
gap between planets c and d is irrelevant. However there is an
additional complication in comparing $\upsilon$ And and 55Cnc. Namely
they may have completely different formation histories. Resonant
systems most likely formed from resonant capture during the migration
epoch (Snellgrove, Papaloizou, \& Nelson 2001), whereas $\upsilon$ And may have formed from a large
scattering event (Rasio \& Ford 1996, Malhotra 2002,
BQ). If a planetary system needs only 2 planets near the edge to be
``packed'', then 55Cnc is already packed, and has no additional companions.

Here we have suggested that additional planets exist in between the known
planets. It could be, though, that additional planets lie beyond
these in an interacting configuration. The work
of Rivera \& Lissuaer (2000) shows that this is the case for the
$\upsilon$ And system (see $\S$4). If the separated systems presented
here actually have more distant companions on interacting orbits,
these systems would still lie close to the edge. Examinations of this
possibility, however, were beyond the scope of this paper.

We have also found that some test particles are more stable on
significantly eccentric orbits. This is true for HD74156 and
HD37124. Although there has been some mention of this phenomenon
(\ie Menou \& Tabachnik 2003), a satisfactory explanation for this
observation is not apparent. Future work should address this issue.

To further explore the PPS scenario, we will integrate the full 4 or 5 body systems with a massive planet in the regions of
stability defined here. This work will be Paper
II. Additionally we will explore the formation of terrestrial planets in these systems in
Paper III. In these papers we will show more evidence that the
predicted planets can survive for upward of 100 million years.

Of course stable regions do not guarantee stability, but it is
nonetheless intriguing to postulate their existence. If the candidate
systems do contain additional companions, then we strengthen
the theory that planet formation is an efficient process and add this
as a new requirement to models of planet formation. Whether or not
this may break the degeneracy between the so-called core accretion
model (\ie Pollack \etal 1996; Bodenheimer, Hubickyj, \& Lissauer 2000) and the gravitational collapse model
(\ie Boss 2002;  Mayer \etal 2002) remains to be seen.

Should the planets predicted here be discovered, it would mark the
first prediction of a planet since Neptune by John Couch Adams in the 19th century. The radial velocity
surveys are providing a rich source of knowledge for the field of
planet formation, altering our understanding of our own solar system,
and changing perspectives of our place in the universe. The correct
prediction of a new planet would represent a major step toward understanding the mechanisms of planet formation.

\section{Acknowledgments}
We thank Thomas Quinn and Beth Willman for useful discussions and suggestions, Chance Reschke for his systems administration during the completion of
these simulations, and the NSF, the NAI, and the NASA GSRP for funding this research. These simulations
were performed on computers donated by the University of Washington
Student Technology Fund, and performed under
CONDOR.\footnote{CONDOR is publicly available at http://www.cs.wisc.edu/condor}

\end{document}